\documentclass[11pt]{article}
\usepackage{amsmath,amsthm,amssymb,cite}
\usepackage{psfrag}
\usepackage{graphicx}
\newcommand{\nc}{\newcommand}
\nc{\mref}[1]{(\ref{#1})}
\nc{\vt}{v_{2\gL_0}}
\nc{\vo}{v_{\gL_0}}
\nc{\vot}{v_{\gL_1+\gL_0}}
\nc{\vw}{v_{\gL_1}}
\nc{\ppmm}{\genfrac{}{}{-10pt}{10pt}{++}{--}}
\nc{\wom}[5]{\Omega\left(\left.\begin{array}{ll}{#1}&{#2}\\{#3}&{#4}\end{array}\right|{#5}\right)}
\nc{\com}[5]{\chi\left(\left.\begin{array}{ll}{#1}&{#2}\\{#3}&{#4}\end{array}\right|{#5}\right)}
\nc{\we}[5]{W\left(\left.\begin{array}{ll}{#1}&{#2}\\{#3}&{#4}\end{array}\right|{#5}\right)}
\nc{\ce}[6]{C^{#6}\left(\left.\begin{array}{ll}{#1}&{#2}\\{#3}&{#4}\end{array}\right|{#5}\right)}
\nc{\lmat}[6]{\ell_{#6}\left(\left.\begin{array}{ll}{#1}&{#2}\\{#3}&{#4}\end{array}\right|{#5}\right)}
\nc{\lmats}[5]{L\left(\left.\begin{array}{ll}{#1}&{#2}\\{#3}&{#4}\end{array}\right|{#5}\right)}
\nc{\hmat}[6]{h_{#6}\left(\left.\begin{array}{ll}{#1}&{#2}\\{#3}&{#4}\end{array}\right|{#5}\right)}
\nc{\hmats}[5]{H\left(\left.\begin{array}{ll}{#1}&{#2}\\{#3}&{#4}\end{array}\right|{#5}\right)}
\nc{\web}[5]{\overline{W}\left(\left.\begin{array}{ll}{#1}&{#2}\\{#3}&{#4}\end{array}\right|{#5}\right)}
\nc{\wep}[5]{W'\left(\left.\begin{array}{ll}{#1}&{#2}\\{#3}&{#4}\end{array}\right|{#5}\right)}
\nc{\wes}[5]{W^*\left(\left.\begin{array}{ll}{#1}&{#2}\\{#3}&{#4}\end{array}\right|{#5}\right)}
\nc{\wess}[5]{W^{**}\left(\left.\begin{array}{ll}{#1}&{#2}\\{#3}&{#4}\end{array}\right|{#5}\right)}
\nc{\cet}[7]{C^{#6}_{#7}\left(\left.\begin{array}{ll}{#1}&{#2}\\{#3}&{#4}\end{array}\right|{#5}\right)}
\nc{\bcet}[7]{\bar{C}^{#6}_{#7}\left(\left.\begin{array}{ll}{#1}&{#2}\\{#3}&{#4}\end{array}\right|{#5}\right)}
\nc{\wet}[7]{W^{#6}_{#7}\left(\left.\begin{array}{ll}{#1}&{#2}\\{#3}&{#4}\end{array}\right|{#5}\right)}
\nc{\bwet}[7]{\overline{W}^{#6}_{#7}\left(\left.\begin{array}{ll}{#1}&{#2}\\{#3}&{#4}\end{array}\right|{#5}\right)}
\nc{\wec}[7]{\widetilde{W}^{#6}_{#7}\left(\left.\begin{array}{ll}{#1}&{#2}\\{#3}&{#4}\end{array}\right|{#5}\right)}
\nc{\wgen}[6]{W^{#6}\left(\left.\begin{array}{ll}{#1}&{#2}\\{#3}&{#4}\end{array}\right|{#5}\right)}
\nc{\wgenp}[6]{W^{*{#6}}\left(\left.\begin{array}{ll}{#1}&{#2}\\{#3}&{#4}\end{array}\right|{#5}\right)}
\nc{\wo}[5]{\Omega\left(\left.\begin{array}{ll}{#1}&{#2}\\{#3}&{#4}\end{array}\right|{#5}\right)}
\nc{\wsgen}[8]{{#8}^{#6}_{#7}\left(\left.\begin{array}{ll}{#1}&{#2}\\{#3}&{#4}\end{array}\right|{#5}\right)}
\nc{\qbinom}[2]{{\genfrac{[}{]}{0pt}{}{{#1}}{{#2}}}_{q}}
\nc{\hhg}[4]{\phi\left({{{#1}\,\,\,{#2}}\atop{{#3}}};
                     {#4}\right)}
\nc{\fullhhg}[5]{ {_2}{\large \phi}_1 \left(   {   { {#1}\,\,\,{#2} } \atop { {#3} } };
                     {#4},{#5}\right)}
\nc{\bra}[1]{\langle #1 |}
\nc{\ket}[1]{| #1 \rangle}
\nc{\qp}[2]{({#1}\, ; \, {#2})_{\infty}}
\nc{\qpf}[1]{({#1}\, ; \, q^4)_{\infty}}
\nc{\pp}[1]{({#1}\, ; \, p)_{\infty}}
\nc{\qpp}[1]{({#1}\, ; \, p, q^4)_{\infty}}
\nc{\sect}{\section}
\nc{\ssect}{\subsection}
\nc{\sssect}{\subsubsection}
\nc{\ud}[1]{\underline{{#1}}}
\nc{\myra}[1]{\buildrel{{#1}}\over \longrightarrow}
\nc{\isomo}{\buildrel {\sim} \over \longrightarrow}
\nc{\Aff}{\operatorname{Aff}}
\nc{\ot}{\otimes}
\nc{\er}{\end{array}}
\nc{\bev}[1]{\begin{equation}\begin{array}{#1}}
\nc{\eeq}{\end{equation}}
\nc{\be}{\begin{eqnarray}}
\nc{\ee}{\end{eqnarray}}
\nc{\ben}{\begin{eqnarray*}}
\nc{\een}{\end{eqnarray*}}
\nc{\bec}{\begin{equation}\begin{array}{lll}}
\nc{\eec}{\end{array}\end{equation}}
\nc{\ed}{\end{document}}
\nc{\half}{\ensuremath{\frac{1}{2}}}
\nc{\Hom}{\operatorname{Hom}}
\nc{\End}{\operatorname{End}}
\nc{\vac}{|\textrm{vac}\rangle}
\nc{\tvac}{|\widetilde{\textrm{vac}}\rangle}
\nc{\dvac}{\langle\textrm{vac}}
\nc{\dtvac}{\langle\widetilde{\textrm{vac}}}
\nc{\id}{\mathbb{I}}
\nc{\ra}{\rightarrow}  
\nc{\lra}{\longrightarrow}
\nc{\uqp}{U^{\prime}_q (\widehat{sl}_2)}
\nc{\uqbp}{U_q (b_+)}
\nc{\uqbm}{U_q (b_-)}
\nc{\ub}{U^{\prime}_q (b_+)}
\nc{\vsl}{V(\sigma(\lambda))}
\nc{\vl}{V(\lambda)}  
\nc{\bu}{\bullet}
\nc{\an}{{\ell}}
\nc{\slth}{\widehat{\mathfrak{sl}}_2\hskip 1pt}

\nc{\ws}{\;\;}
\nc{\qu}{{1\ov 4}}
\nc{\hif}{\hb{ if }}
\nc{\hev}{\hb{ is even }}
\nc{\hod}{\hb{ is odd }}
\nc{\Tr}{{\rm Tr}}
\nc{\ad}{{\rm Ad}}
\nc{\hb}{\hbox}
\nc{\nn}{\nonumber} 
\nc{\curlra}{\buildrel{\sim}\over\longrightarrow}
\nc{\epp}{\varepsilon^{\prime}} 
\nc{\ol}{\overline}
\nc{\pl}{\prod\limits} 
\nc{\sli}{\sum\limits} 
\nc{\nin}{\noindent}

\nc{\ga}{\alpha}
\nc{\gb}{\beta}
\nc{\gd}{\delta}
\nc{\gep}{\varepsilon}
\nc{\gz}{\zeta}
\nc{\gt}{\theta}
\nc{\gk}{\kappa}
\nc{\gl}{\lambda}
\nc{\gp}{\phi}
\nc{\gs}{\sigma}
\nc{\go}{\omega}
\nc{\gn}{\nu}
\nc{\gr}{\rho}

\nc{\gou}{\underline{\go}}
\nc{\un}{\underline{n}}
\nc{\um}{\underline{m}}
\nc{\uw}{\underline{w}}

\nc{\s}{\sigma}
\nc{\ep}{\varepsilon}
\nc{\z}{\zeta}
\nc{\g}{\gamma}
\nc{\zi}{\zeta^{-1}}

\nc{\gG}{\Gamma}
\nc{\gD}{\Delta}
\nc{\gT}{\Theta}
\nc{\gL}{\Lambda}
\nc{\gO}{\Omega}
\nc{\gP}{\Phi}

\nc{\cL}{\mathcal{L}}
\nc{\cF}{\mathcal{F}}
\nc{\cP}{\mathcal{P}}
\nc{\cS}{\mathcal{S}}
\nc{\cN}{\mathcal{N}}
\nc{\cD}{\mathcal{D}}
\nc{\cH}{\mathcal{H}}
\nc{\cO}{\mathcal{O}}
\nc{\cT}{\mathcal{T}}
\nc{\cQ}{\mathcal{Q}}
\nc{\cW}{\mathcal{W}}
\nc{\cR}{\mathcal{R}}

\nc{\C}{\mathbb{C}}
\nc{\Q}{\mathbb{Q}}
\nc{\R}{\mathbb{R}}
\nc{\Z}{\mathbb{Z}}
\nc{\N}{\mathbb{N}}

\nc{\fg}{\mathfrak{g}}

\nc{\barx}{\bar{x}}
\nc{\bi}{\bar{i}}
\nc{\bj}{\bar{j}}
\nc{\bgr}{\bar{\rho}}
\nc{\bA}{\bar{\alpha}}
\nc{\bB}{\bar{\beta}}
\nc{\bC}{\bar{\gamma}}
\nc{\by}{\bar{y}}
\nc{\brv}{\overline{V}}
\nc{\brp}{\overline{P}}

\nc{\tf}{\tilde{f}}
\nc{\te}{\tilde{e}}
\nc{\ts}{\tilde{s}}
\nc{\tgP}{\widetilde{\Phi}}
\nc{\tgPs}{\tilde{\Psi}}
\nc{\tgn}{\tilde{\nu}}
\nc{\tgl}{\tilde{\lambda}}
\nc{\tge}{\tilde{\eta}}
\nc{\txi}{\tilde{\xi}}
\nc{\tep}{\tilde{\epsilon}}

\nc{\cB}{\check{b}}
\nc{\cOm}{\check{\Omega}}

\nc{\goto}{\mapsto}
\nc{\embed}{\hookrightarrow}
\nc{\rien}{\emptyset}
\nc{\lb}[1]{\label{#1}}
\nc{\Nt}{\frac{N}{2}}
\nc{\vn}{\hspace*{-33truemm}}
\nc{\vm}{\hspace*{-0truemm}}
\nc{\ti}{t^{-1}}
\nc{\vb}{v^{(1)}}
\nc{\vbn}{v^{(n)}}
\nc{\ur}{\underline{r}}
\nc{\us}{\underline{s}}
\nc{\up}{\underline{p}}
\nc{\bp}{\bar{p}}
\nc{\bpi}{\bar{p}^{(i)}}
\nc{\bpip}{\bar{p}^{(i+1)}}
\nc{\vz}{V^{(1)}_z}
\nc{\vzn}{V^{(n)}_z}
\nc{\vzo}{V^{(1)}_1}
\nc{\piz}{\pi_z^{(1)}}
\nc{\pizn}{\pi_z^{(n)}}
\nc{\pis}{\pi_{(z,\us)}}
\nc{\bW}{\overline{W}}
\nc{\bQ}{\overline{Q}}
\nc{\tQ}{\widetilde{Q}}
\nc{\bT}{\overline{T}}
\nc{\note}[1]{\vspace*{-5mm}\marginpar[left]{\scriptsize\bf{#1}}}
\nc{\mynote}[1]{\nin{\bf Note:{#1}}\newline}
\nc{\eqdef}{:=}
\nc{\lu}{^{(\lambda)}}
\nc{\vone}{v_{\Lambda_1}}
\nc{\vzero}{v_{\Lambda_0}}

\def\bea{\begin{eqnarray}}
\def\eea{\end{eqnarray}}

\def\ep{\varepsilon}

\newtheorem{theorem}{Theorem}[section]

\newtheorem{example}[theorem]{Example}
\def\by{{\mbox{\boldmath $y$}}}

\def\bpm{\begin{pmatrix}}
\def\epm{\end{pmatrix}}
\def\bdet{\left|\begin{array}}
\def\edet{\end{array}\right|}
\def\ds{\displaystyle}
\nc{\dx}[1]{\frac{d{#1}}{dx}}
\nc{\ddx}[1]{\frac{d^2{#1}}{dx^2}}
\nc{\dt}[1]{\frac{d{#1}}{dt}}
\nc{\ddt}[1]{\frac{d^2{#1}}{dt^2}}

\newcounter{quest}
\newlength{\questoffset}
\nc{\pd}{\partial}
\nc{\nl}{\newline}
\nc{\vp}{\vspace*{3mm}}
\nc{\soln}{\noindent {\it Solution}:\,}
\nc{\pro}{\noindent {\it Proof}\,:\,}
\nc{\bex}{\begin{example}}
\nc{\eex}{\end{example}}

\nc{\mitem}{\noindent}

\newcommand{\ena}{\end{eqnarray}}

\def\cip(#1){(#1;p,q^4)_\infty}
\def\z{\zeta}

\def\Lpm#1{\mathrel{\mathop{\kern0pt L^\pm}\limits^#1}}
\def\L#1{\mathrel{\mathop{\kern0pt L}\limits^#1}}
\def\Lm#1{\mathrel{\mathop{\kern0pt L^-}\limits^#1}}
\def\Lp#1{\mathrel{\mathop{\kern0pt L^+}\limits^#1}}
\def\vep{\varepsilon}

\newcommand\tx{\tilde{x}}
\begin{document}
\bibliographystyle{unsrt}
\begin{flushright}
\end{flushright}
\begin{center}
{\LARGE \bf Exact and Scaling Form of the Bipartite Fidelity of the Infinite XXZ Chain}\\[10mm]
{\Large \bf 
Robert Weston}\\[3mm]
{\it Department of Mathematics, Heriot-Watt University,\\
Edinburgh EH14 4AS, UK.
}\\[5mm]
March 2012\\[10mm]

\end{center}
\begin{abstract}
\noindent 
We find an exact expression for the bipartite fidelity $f=|\langle \hb{vac} |\hb{vac}\rangle'|^2$, where $|\hb{vac}\rangle$ is the vacuum eigenstate of an infinite-size antiferromagnetic XXZ chain and $|\hb{vac}\rangle'$ is the vacuum eigenstate of an infinite-size XXZ chain which is split in two. We consider the quantity $-\ln(f)$ which has been put forward as a measure of quantum entanglement,  and show that the large correlation length $\xi$ behaviour is consistent with a general conjecture 
$-\ds\ln(f)\sim \frac{c}{8}\ln(\xi)$, where $c$ is the central charge of the $UV$ conformal field theory (with $c=1$ for the XXZ chain). This behaviour is a natural extension of the existing conformal field theory prediction of $\ds-\ln(f)\sim \frac{c}{8}\ln(L)$ for a length $L$ bipartite system with $0\ll L\ll\xi$.
\end{abstract}
\nopagebreak

\section{Introduction}
In this letter, we consider two separate Hamiltonians: the usual infinite-size XXZ Hamiltonian $H$, and the infinite-size XXZ Hamiltonian $H'$ with the interaction between central adjacent sites $0$ and $1$ removed. The precise definitions of these Hamiltonians are given by \mref{eqn:ham1} and \mref{eqn:ham2} below. Specialising the previous results of the author \cite{RW2011}, we obtain exact expressions for the respective vacua $|\hb{vac}\rangle$ and $|\hb{vac}\rangle'$ and for the bipartite fidelity $f=|\langle \hb{vac}| \hb{vac}\rangle'|^2$.

 The quantity $-\ln(f)$ has been put forward as a measure of quantum entanglement in \cite{Dubail2011}. This interpretation is physically reasonable: the closer $|\hb{vac}\rangle$ is to a pure tensor-product state, the closer $f$ will be to 1 and the smaller will be $-\ln(f)$.
Another way in which $-\ln(f)$ resembles other measures of quantum entanglement, such as the standard von Neumann entanglement entropy, is in its universal scaling behaviour. The scaling behaviour of $-\ln(f)$ has been considered in the setting of conformal field theory in \cite{Dubail2011}. Using the results of Cardy and Peschel for the free energy of a conformal field theory on a 2D sheet of size $L\times L$ with a semi-infinite slit \cite{CardyPeschel88}, Dubai and St\'ephan derived the following universal form for $-\ln(f)$ for a length $L$ 1D bipartite quantum system \cite{Dubail2011}:
\bea -\ln(f) ~\substack{\\\sim\\ L \rightarrow \infty }~  \frac{c}{8}\,\ln(L),\lb{eqn:fcft}\eea
where $c$ is the central charge of the UV conformal field theory. 
This result is valid in the case when the correlation length $\xi$ is much larger than the system size $L$.  In this letter, we consider the alternative regime in which $\xi$ is finite and $L$ infinite. We find the exact expression for $f$ given by \mref{eqn:f}, and the scaling behaviour 
\bea -\ln(f)  ~\substack{\\\sim\\ \xi \rightarrow \infty }~     \frac{c}{8}\, \ln(\xi),\lb{eqn:lfconj}\eea
with the constant $c$ equal to 1 for the antiferromagnetic XXZ model. We conjecture that when $0\ll \xi\ll L$ the universal form \mref{eqn:lfconj} should hold for any 1D quantum system which is a perturbation of a UV conformal field theory with central charge $c$ (and which has a trivial $c=0$ IR fixed point).

It is interesting to note that the entanglement entropy $-\Tr(\rho \ln \rho)$ of a 1D bipartite system (in which $\rho$ is the reduced density matrix associated with the half-line) has a universal scaling behaviour of the form \mref{eqn:lfconj} but with the coefficient $\ds\frac{c}{8}$ replaced by  $\ds\frac{c}{6}$ \cite{Hol94,CardyCalabrese2004}. A general field theory argument leading to this latter result for the entanglement entropy and exact calculations for the XY and XXZ lattice models are given in \cite{Peschel2004,CardyCalabrese2004,RW2006}. 
Apart from the the different $c$ coefficient, the scaling form of the bipartite fidelity we find for the XXZ model differs
 from the analogous expression for the entanglement entropy in another important respect: there are no finite correction terms as $\xi\ra\infty$ (see Expression \mref{eqn:lnf}). Such corrections {\it are} present for the entanglement entropy and 
in that context they have been interpreted as boundary entropy contributions \cite{CardyCalabrese2004,RW2006}.

\section{The Model}
In this section we define our two different quantum spin chain Hamiltonians. The first is the infinite-size antiferromagnetic XXZ Hamiltonian
\bea H=-\half \sli_{j\in \Z} \left (\sigma_{j}^x \sigma_{j+1}^x + \sigma_{j}^y \sigma_{j+1}^y+\Delta 
\sigma_{j}^z \sigma_{j+1}^z\right),\quad
\Delta=-\frac{(x+x^{-1})}{2}, \ws 0<x<1.\lb{eqn:ham1}\eea
This Hamiltonian acts on the space with antiferromagnetic boundary conditions $+-+-+-$ at plus and minus infinity (we use $+$ and $-$ to refer to the two eigenstates of $\sigma^z$). The second Hamiltonian $H'$ is split into $H'=H_L+H_R$, where $H_L$  acts on a left-hand semi-infinite space and $H_R$ acts on a right-hand semi-infinite space with the same boundary conditions as above. We have
\bea
H'&=&H_L+H_R,\quad \hb{with}\nn\\
H_L&=& -\half \sli_{j\leq -1} \left (\sigma_{j}^x \sigma_{j+1}^x + \sigma_{j}^y \sigma_{j+1}^y+\Delta \sigma_{j}^z \sigma_{j+1}^z\right),\lb{eqn:ham2}\\
H_R&=& -\half \sli_{j\geq 1} \left (\sigma_{j}^x \sigma_{j+1}^x + \sigma_{j}^y \sigma_{j+1}^y+\Delta \sigma_{j}^z \sigma_{j+1}^z\right).\nn\eea
Note that $H'$ differs from $H$ only in the absence of any interaction between the sites at positions $0$ and $1$. Exact expressions for the vacuum eigenstates $|\hb{vac}\rangle$ and $|\hb{vac}\rangle'$ of $H$ and $H'$ and all correlation functions $\langle \hb{vac}| {\cal O} |\hb{vac}\rangle'$ were obtained by the author in \cite{RW2011} by exploiting the vertex operator approach to bulk and boundary quantum spin chains \cite{JM,JKKKM,JKKMW}. In fact, a slightly more general split Hamiltonian $H'+h(\sigma_0^z-\sigma_1^z)$ was considered in \cite{RW2011}, but in the current paper we set the magnetic field $h=0$ corresponding to the choice of $r=-1$ in the notation of \cite{RW2011}.

\section{The Exact Bipartite Fidelity}
The bipartite fidelity $f=|\langle \hb{vac} |\hb{vac}\rangle'|^2$ can be read off from Equation (4.1) of \cite{RW2011} with the specialisation $r=-1$. Using the infinite-product notation 
\bea (z;a_1,a_2,\cdots,a_N)_\infty=\prod\limits_{n_1=0}^\infty\prod\limits_{n_2=0}^\infty\cdots\prod\limits_{n_N=0}^\infty(1-z\, a_1^{n1} a_2^{n_2}\cdots a_N^{n_N})\lb{eqn:infprod}\eea
 we have the raw result
\ben f=(x^2;x^4)_\infty \frac{(x^{6};x^8,x^8)^2_\infty \,(x^{10};x^8,x^8)^2_\infty}{(x^{4};x^8,x^8)^2_\infty \, (x^{12};x^8,x^8)^2_\infty}\frac{(x^{2};x^4,x^8)^2_\infty}{(x^{4};x^4,x^8)^2_\infty}.\een
This expression simplifies considerably if we make use of the q-calculus identities 
\bea (z;x^{2b},x^c)_\infty (zx^b; x^{2b},x^c)_\infty&=&  (z;x^{b},x^c)_\infty,\lb{eqn:qcalc1} \\ (z;x^{b},x^c)_\infty (-z; x^{b},x^c)_\infty&=&  (z^2;x^{2b},x^{2c})_\infty, \lb{eqn:qcalc2}\eea
which follow easily from the infinite-product definition of Equation \mref{eqn:infprod} (an introduction to q-calculus can be found for example in the chapter by G.E. Andrews in \cite{NIST}). Using the identities \mref{eqn:qcalc1} and \mref{eqn:qcalc2} successively we find
\bea f=(x^2;x^4)_\infty \frac{(-x^4;x^4,x^4)^2_\infty}{(-x^2;x^4,x^4)^2_\infty}.\lb{eqn:f}\eea

\section{The Large Correlation Length Behaviour}
The XXZ model is exactly solvable and the correlation length $\xi$ of the antiferromagnetic XXZ Hamiltonian $H$ is given by \cite{JKM73,Bax82}
\ben \xi^{-1}=-\half \ln(k(x^2))=-\half \ln\left(\frac{1-k'(x)}{1+k'(x)}\right),\een
where $k$ and $k'$ are the the elliptic modulus and dual modulus functions\footnote{The basic properties of elliptic functions and their modular transformations are described in many places including Chapter 15 of \cite{Bax82} and \cite{NIST}.}
 given by
\bea k(z)=4 \,z^\half \frac{(-z^2;z^2)^4_\infty}{(-z;z^2)^4_\infty},\quad k'(z)=\frac{(z;z^2)_\infty^4}{(-z;z^2)_\infty^4}.
\lb{eqn:kdef}\eea
The $\xi\ra \infty$ limit corresponds to $x\ra 1$ (or equivalently $\Delta \ra -1$) at which point $k'(x)\ra 0$ and we have\\[-11mm]
\ben \ln(\xi) ~\substack{\\=\\ k' \rightarrow 0 }~ -\ln(k')+ O(k'^2).\een
If we parametrise $x=e^{-\ep}$, and define $\tx=e^{-\pi^2/\vep}$, then we have $k'(x)=k(\tx)$  \cite{Bax82}. The behaviour of $\ln(k(\tx))$ as $\tx\ra 0$ is then specified by \mref{eqn:kdef}, which leads to\\[-5mm] 
\ben \ln(\xi) ~\substack{\\=\\ \vep \rightarrow 0 }~ \frac{\pi^2}{2\vep}-\ln(4) + O(\vep).\label{eqn:lnxi}\een

The $\vep\ra 0$ behaviour of $-\ln(f)$ may be computed from \mref{eqn:f} using the method described in Appendix A which again relies primarily on the known $x\ra \tx$ modular transformation properties of elliptic functions. 
We find 
\bea -\ln(f)~\substack{\\=\\ \vep \rightarrow 0 }~\frac{\pi^2}{16\vep}-\frac{1}{4}\ln(2) + O(\vep),\lb{eqn:id1}\eea
and hence arrive at the result
\bea -\ln(f)~\substack{\\=\\ \vep \rightarrow 0 }~ \frac{1}{8} \ln(\xi)+O(\vep).\lb{eqn:lnf}\eea

The antiferromagnetic XXZ model has a UV fixed point described by a $c=1$ conformal field theory (namely the $\slth$ WZW model at level 1 \cite{FMS}). 
Hence, by analogy with the conformal field theory prediction \mref{eqn:fcft}, we are led to the following conjecture for the universal scaling behaviour of a 1D bipartite system with $0\ll\xi\ll L$ which is a perturbation of a UV conformal field theory with central charge $c$ and which has a trivial $c=0$ IR fixed point:
\bea -\ln(f)  ~\substack{\\\sim\\ \xi \rightarrow \infty }~     \frac{c}{8}\, \ln(\xi).\lb{eqn:lfconj2}\eea

\section{Comments}
In this letter, we have used the language of 1D quantum systems. There is of course an interpretation of $-\ln(f)$ in terms of the free energy of a 2D classical statistical-mechanical system - the 6-vertex model. This `fractured' 6-vertex model with a semi-infinite slit from the centre was discussed in detail in \cite{RW2011}. In particular, the boundary conditions along the slit, the construction of the partition function and correlation function in terms of corner and semi-infinite transfer matrices, and an exact expression for all correlation functions  $\langle \hb{vac}| {\cal O} |\hb{vac}\rangle'$ can be found in \cite{RW2011}.

Two obvious questions arise in association with our conjecture \mref{eqn:lfconj2}: does it indeed hold for other examples of exactly-solvable 1D lattice models, and can it be proved using field theory arguments along the lines of those deployed in the analysis of the scaling behaviour of the entanglement entropy with $0\ll\xi\ll L$ 
\cite{CardyCalabrese2004,CardyCastro2008,Doyon2008}? Both questions will be addressed by the author in future publications. 
\newpage

\begin{appendix}
\section{ Asymptotic behaviour of $\ln(f)$}
In this appendix we derive the $x \ra 1$ behaviour of the log of the fidelity $f$ given by Equation \mref{eqn:f}. Our strategy is to rewrite $f$ in terms of a function with simple behaviour under the modular transformation $x=e^{-\vep}\ra \tx= e^{-\pi^2/\vep}$ times a function that  is manifestly convergent when $x \ra 1$. To this end, we use the q-calculus identity 
\ben (-x^4;x^4,x^4)_\infty=\frac{(-1;x^4,x^4)_\infty}{2(-x^4;x^4)_\infty}\een to obtain\\[-9mm] 
\bea f&=&\frac{(x^2;x^4)_\infty}{2(-x^4;x^4)_\infty }\, g,\lb{eqn:fg}\\
\hb{where}\quad g&=&\frac{(-1;x^4,x^4)_\infty (-x^4;x^4,x^4)_\infty}{(-x^2;x^4,x^4)^2_\infty}.\lb{eqn:gdef}\eea
The non-$g$ terms in \mref{eqn:fg} may be rewritten using the identity 
\bea x^{-\frac{b}{48}}\, (x^\frac{b}{2};x^b)_\infty =\sqrt{2}\, \tx^{\frac{1}{6b}} \,(-\tx^\frac{4}{b};\tx^\frac{4}{b})_\infty,\lb{niceid}\eea
which may be derived either by identifying $(x^\frac{b}{2};x^b)^2_\infty$ as a short theta function and using 
the modular transformation properties of the latter\footnote{A clear description of the definition and properties of short theta 
functions is given in \cite{FV99}.}, or more directly by Poisson resummation of the log of both sides \cite{Bax82}. Making use of \mref{niceid} allows us to rewrite 
\bea f =x^\frac{1}{4} \,\tx^\frac{1}{16} \, \frac{(-\tx;\tx)_\infty}{(\tx^\half;\tx)_\infty }\, g .\lb{eqn:ftrans}\eea

The log of the function $g$ given by \mref{eqn:gdef} may be expressed by using \ben
\ln\big( (-z;x^4,x^4)_\infty \big)= \sli_{N= 1}^\infty \frac{ (-1)^{N+1} }{ N } \frac{z^N}{(1-x^{4N})^2},\een
from which we obtain
\ben \ln(g)=\sli_{N=1}^\infty \frac{(-1)^{N+1}}{N} \frac{1}{(1+x^{2N})^2} .\een
This is convergent as $x=e^{-\vep} \ra 1$ with
\ben \ln(g)~\substack{\\=\\ \vep \rightarrow 0 }~ \frac{1}{4} \ln(2)+O(\vep).\een
The $x=e^{-\vep} \ra 1$, $\tx=e^{-\pi^2/\vep}\ra 0$ behaviour of the non-$g$ terms in \mref{eqn:ftrans} is also well defined and is controlled by the $\tx^{1/16}$ term. We thus obtain 
\bea \ln(f) ~\substack{\\=\\ \vep \rightarrow 0 }~ -\frac{\pi^2}{16 \vep} + \frac{1}{4} \ln(2)+O(\vep).\lb{eqn:lnfscaling}\eea

\end{appendix}

\end{document}